\documentstyle[multicol,aps,prl,epsf,epsfig,psfig]{revtex}
\begin{document}
\title{Slow Logarithmic Decay of Magnetization in the Zero Temperature Dynamics of an Ising
Spin Chain: Analogy to Granular Compaction} \author{Satya
N. Majumdar$^{1,2}$, David S. Dean$^1$ and Peter Grassberger$^3$\\ {\small 1. Laboratoire de
Physique Quantique (UMR C5626 du CNRS), Universit\'e Paul Sabatier,
31062 Toulouse Cedex, France}\\ {\small 2. Tata Institute of
Fundamental Research, Homi Bhabha Road, Mumbai-400005, India}\\
{\small 3. NIC, Forschungszentrum J\"ulich, D-52425 J\"ulich, Germany}}

\date{april 22 2000}
\maketitle

\begin{abstract}
We study the zero temperature coarsening dynamics in an Ising chain in
presence of a dynamically induced field that favors locally the `$-$'
phase compared to the `$+$' phase. At late times, while the `$+$' domains 
still coarsen as $t^{1/2}$, the `$-$' domains coarsen slightly faster as $t^{1/2}\log (t)$.
As a result, at late times, the magnetization decays 
slowly as, $m(t)=-1 +{\rm const.}/{\log (t)}$. We establish this behavior both
analytically within an independent interval
approximation (IIA) and numerically. In the zero volume fraction limit
of the `$+$' phase, we argue that the IIA becomes asymptotically exact. 
Our model can be alternately viewed as a simple Ising model for granular compaction.
At late times in our model, the system decays into a fully compact state (where all spins are `$-$')
in a slow logarithmic manner $\sim 1/{\log (t)}$, a fact that has been observed in recent experiments on
granular systems.

\vskip 0.5cm


\end{abstract}

\begin{multicols}{2}
The effect of {\it quenched disorder} on the relaxation dynamics of
many body systems has been studied quite extensively\cite{BCKM}. In systems
such as structural glasses, where quenched disorder is absent, an
alternative approach has been put forward that considers the slow
relaxation due to {\it kinetic disorder}, induced by the dynamics
itself\cite{Fred}. Another important system where kinetic disorders
give rise to slow relaxation is granular material. The density relaxation of loosely 
packed glass beads has been studied in recent experiments and it
was found that the density $\rho(t)$ compactified slowly as, $\rho (\infty)-\rho (t)\sim 1/{\log (t)}$,
under mechanical tapping\cite{GE}.    
It is natural to expect that such kinetic
disorders may play important roles in the dynamics of other systems
as well. In this Letter we study, for the first time, the effect of
a dynamically self-induced {\it field} in an important class of out of
equilibrium problems, namely the domain growth problems, and show that
such systems also exhibit logarithmic relaxation, suggesting that
inverse logarithmic relaxation is quite robust.

Domain growth following a rapid quench in temperature in ferromagnetic
spin systems is one of the better understood out of equilibrium
phenomena\cite{bray}.  For example, if an Ising system is quenched
rapidly from infinite temperature to zero temperature without breaking
the symmetry between the two ground states, domains of up and down
spins form and grow with time. The average linear size of a domain
grows with time as $l(t) \sim t^{1\over 2}$ for zero temperature
nonconserved dissipative dynamics. However if one puts on a small
uniform external field (say in the down direction), then even at zero
temperature the system quickly reaches the pure state of magnetization
$-1$ in a finite time proportional to the initial size of the
up domains.  A natural question is: what happens when, instead of
applying a global external bias, the symmetry between the pure states
is broken locally by the dynamics itself ?

In this Letter we address this question in the context of a simple
Ising spin chain with spins $S_i = \pm 1$.  Starting from a given
initial configuration, the system evolves by single spin flip
continuous time dynamics.  Let $W(S_i;S_{i-1}, S_{i+1})$ denote the
rate at which the flip $S_i \to -S_i$ occurs when the two neighboring
spins are $S_{i-1}$ and $S_{i+1}$.  In our model the rates are
specified as follows:
\begin{eqnarray}
& & W(+;++) = W(-;--) = 0 \nonumber \\ & & W(+;-+) = W(+;-+) = W(-;+-)
= W(-;-+) = {1\over 2} \nonumber\\ & & W(+;--) = 1 \nonumber\\ & &
W(-;++) = \alpha
\label{eq:rates}
\end{eqnarray}
Here we restrict ourselves to the case $\alpha = 0$. We note that the
case $\alpha = 1$ corresponds to the usual zero temperature Glauber
dynamics\cite{glauber}.  The only difference is that for $\alpha=0$,
the move $(+,-,+) \to (+,+,+)$ is not allowed and thereby the symmetry
between `$+$' and `$-$' spins is locally dynamically broken. Thus
isolated `$-$' spins (surrounded on both sides by a `$+$') block the
coalescence of `$+$' domains and locally favor the `$-$' spins. As
a result, we show below, the system eventually decays into the state
where all spins are `$-$' but does so in a very slow manner, $m(t)+1 \sim 1/{\log (t)}$.

Our main results can be summarized as follows. In contrast to 
the case $\alpha = 1$ (where
the average size of both `$+$' and `$-$' domains grow as
$\l_{\pm}(t)\sim t^{1/2}$ at late times\cite{bray1} and the average
magnetization $m(t) = (l_+ - l_-)/(l_+ + l_-)$ is a constant of
motion\cite{glauber}), for $\alpha = 0$ we show that at late times,
while $l_{+}(t)\sim t^{1/2}$, $l_{-}(t)\sim t^{1/2}\log (bt)$ 
where $b$ is a constant that depends on the initial condition.
Thus due to the dynamically generated local bias, the `$-$' domains grow
slightly faster than the `$+$' domains and as a result the
magnetization decays as, $m(t) = -1 + {\rm const}/{\log (bt)}$ for large
$t$. Notice that the average domain size grows faster for $\alpha = 0$ 
than for $\alpha = 1$, i.e. paradoxically coarsening is {\it enhanced} 
by putting one of the rates to zero.

Our model can alternately be viewed as a toy model of granular compaction
if one identifies the `$-$' spins as particles,`$+$' spins as holes and
the $1$-d lattice as a section of the bottom layer of a granular system.
The final state where all spins are negative (magnetization, $m=-1$) then
corresponds to the fully compact state with particle density $1$. The basic 
mechanism for granular compaction can be summarized as follows. There exist
local kinetic `defects' and the system can gain in compaction only by relaxing such 
local defects. Such relaxation happens via the tapping process. However, these
defects become rarer with time and it becomes harder and harder for the system
to find such a local defect, relax it and thereby gain in compaction. This is the
origin of the slow logarithmic relaxation. In our model, the triplets `$+-+$' play the role
of such local defects which decay only via the diffusion of kinks. Thus the diffusion
effectively plays the role of tapping. The density of these triplets decays with
time and the system finds it progressively harder to relax. Such logarithmic relaxation
has been found previously in various models of granular systems\cite{GM}. However our model
differs from these models in an important way. In previous models, the density, while increasing
as $\rho (\infty)-\rho (t)\sim 1/{\log (t)}$ at intermediate times, finally
saturates to its steady state value
exponentially fast\cite{GM} for finite values of the rates of microscopic processes. 
In contrast, in our model,
the approach to the final
state is logarithmic asymptotically even at very late times.  \\

In terms of the motion of the domain walls between `$+$' and `$-$'
phases, our model can also be viewed as a reaction diffusion process. 
We need to distinguish between the two types
of domain walls $-+ \equiv A$ and $+- \equiv B$. Note that by
definition (originating from a spin configuration) the $A$'s and $B$'s
always occur alternately. The $A$'s and $B$'s diffuse and
when an $A$ and a $B$ meet, they annihilate only if $A$ is
to the left of $B$, otherwise there is hard core repulsion between
them.

To start with, we set up our notations.  
We define $P_n(t)$ and $R_n(t)$ to be the number of domains of size $n$ per unit
length of `$+$' and `$-$' types respectively. Then, $N(t)=\sum_n P_n=\sum_n R_n$ is the number of domains
of either `$+$' or `$-$' spins per unit length. The density of kinks is therefore $2N(t)$.
We also define the normalized variables, $p_n=P_n/N$ and $r_n=R_n/N$. $p_n$ (or $r_n$) denotes
the conditional probability that given a domain of `$+$' (or `$-$') has occurred, it is of length $n$.
Let $L_+(t)=\sum nP_n$ and $L_-(t)=\sum nR_n$ denote the densities of
`$+$' and `$-$' spins. Clearly $L_+(t)+L_-(t)=1$ and  
the magnetization per unit length is $m(t)=L_+(t)-L_-(t)$.
The average size of a 
`$+$' and a `$-$' domain are denoted respectively by $l_+(t)=\sum np_n=L_+(t)/N$ and $l_-(t)=\sum nr_n=L_-(t)/N$.

Following Glauber's calculation for the $\alpha=1$ case, it is easy to show\cite{MDG} that for $\alpha=0$ case, 
the domain density $N(t)=(1-\langle S_iS_{i+1}\rangle)/4$ of either phase, and the fraction of `$+$' spins,
$L_+(t)=(1+\langle S_i\rangle)/2$ evolve according to the exact equations,
\begin{equation}
{dN\over dt} = -P_1 ,
\label{exactP:N}
\end{equation}        
and
\begin{equation}
{dL_+\over dt} = - R_1.
\label{exactR:L}
\end{equation}
where $P_1(t)=  \langle (1 - S_{i-1})(1 +S_{i}) (1 - S_{i+1})
\rangle/8$ and $R_1(t) = \langle (1 + S_{i-1})(1 - S_{i}) (1 + S_{i+1})
\rangle/8 $ are respectively the density of `$+$' and `$-$' domains of unit length, i.e., the density of triplets
`$-+-$' and `$+-+$. It is easy to see physically the origin of these two exact equations.
Eq.(\ref{exactP:N}) arises from the fact that the domain density
can decrease only via the annihilation of the triplets `$-+-$'.
Also, on average, the fraction of `$+$' spins can decrease only due to the blockage
by `$+-+$' triplets giving rise to Eq.(\ref{exactR:L}).
                                                          
Using $m(t)=\langle S_i\rangle =2L_+(t)-1$, we find from Eq.(\ref{exactR:L}) that $dm/dt=-2R_1$.
We note that for the case $\alpha = 1$, $dm/dt=0$ \cite{glauber}, indicating that the magnetization
does not evolve with time. In our case, due to the triplet defects
`$+-+$', the average magnetization decays with time.  
We also note that, unlike the $\alpha = 1$ case, the
evolution equation (\ref{exactR:L}) for the single point correlation
function involves two and three point correlations (via $R_1(t)$). 
Writing down the analogous equations for $R_n(t)$ gives an infinite 
hierarchy which makes an exact solution difficult for  $\alpha =0$.
  
Using $R_1=r_1 N$ and $L_+=l_+ N$ in Eq.(\ref{exactR:L}), one can formally solve for $N(t)$ in terms of $r_1$ and
$l_+$ as,
\begin{equation}
{N(t) \over N(t_0)} = {l_+(t_0)\over l_+(t)}\exp\left(-\int_{t_0}^t
{r_1(t')\over l_+(t')}dt' \right).
\label{Nsol1}
\end{equation}    
Furthermore if the density of the `$+$' phase is $L_{+}(t_0)=\epsilon$, then, using the relation
$N(t) = 1/[l_{-}(t) + l_{+}(t)]$ in Eq.(\ref{Nsol1}), we find
\begin{equation}
{l_-(t)\over l_+(t)} = {1\over \epsilon}\exp\left(\int_{t_0}^t
{r_1(t')\over l_+(t')}dt' \right) - 1,
\label{ratio1}
\end{equation}
clearly showing that the ratio $l_-(t)/l_+(t)$ is growing due to the presence of the
triplets `$+-+$'. Note that the asymmetry between the growth of `$-$' and `$+$' domains
is evident due to the triplet defects `$+-+$', present with density $R_1=r_1 N$. 
 
In order to make further analytic progress, we first use the IIA where
correlations
between neighboring domains are neglected. The IIA was used previously
for the pure Glauber-Ising model, i.e., the $\alpha=1$ case\cite{KB}. It yielded results in agreement,
qualitatively as well as quantitatively to a fair degree of accuracy,
with the exact results available \cite{glauber,DZ}.  Following the derivation of the IIA equations in the $\alpha=1$
case, it is straightforward to derive the corresponding equations for the $\alpha=0$ case\cite{MDG}. Under this
approximation, the domain densities $P_n(t)$ and
$R_n(t)$ evolve as\cite{MDG}
\begin{equation}
{dP_n\over dt} = P_{n+1} + P_{n-1} - 2 P_n + {R_1\over N}(P_n - P_{n-1})
\label{eq:P_n}
\end{equation}
for all $n\geq 1$ with $P_0=0$ (absorbing boundary condition) and
\begin{eqnarray}
{dR_n\over dt} &=& R_{n+1} + R_{n-1} - 2 R_n -{P_1\over N}(R_n+R_{n-1})  \nonumber \\
&\ & \ \ \ \ + {P_1\over N^2} \sum_{i=1}^{n-2}R_i R_{n-i-1} ; \ \ \ n\geq 2
\nonumber \\ {dR_1\over dt}&=& R_2-R_1 -{P_1\over N}R_1 ,
\label{eq:R_n}
\end{eqnarray}
where $N(t)=\sum P_n=\sum R_n$. It can be easily checked that these two IIA equations satisfy
Eqs.(\ref{exactP:N}) and (\ref{exactR:L})
exactly, and consequently also Eqs.(\ref{Nsol1}) and (\ref{ratio1}).

To calculate $N(t)$ using Eq.(\ref{Nsol1}),
we need to evaluate two quantities from the IIA equations: (i) $r_1(t)=R_1/N$ and (ii) $l_+(t)=\sum np_n$ where
$p_n=P_n/N$.
In order to calculate these two quantities, it is useful to write the IIA equations in terms of the
normalized variables, $p_n=P_n/N$ and $r_n=R_n/N$. From Eqs.(\ref{eq:P_n}) and (\ref{eq:R_n}), we then get,
\begin{equation}
{dp_n\over dt} = p_{n+1} + p_{n-1} - 2 p_n + r_1(p_n - p_{n-1}) + p_1
p_n
\label{eq:p_n}
\end{equation}
for all $n\geq 1$ with $p_0=0$ (absorbing boundary condition) and
\begin{eqnarray}
{dr_n\over dt} &=& r_{n+1} + r_{n-1} - 2 r_n -p_1 r_{n-1} \nonumber \\
&\ & \ \ \ \ + p_1 \sum_{i=1}^{n-2}r_i r_{n-i-1} ; \ \ \ n\geq 2
\nonumber \\ {dr_1\over dt}&=& r_2-r_1 .
\label{eq:r_n}
\end{eqnarray}
It is easy to check that the normalization condition $\sum p_n=\sum r_n =1$ is satisfied by these two equations.

The two IIA equations above are coupled nonlinear equations with infinite number of variables and
hence are difficult to solve exactly. Our approach is a combination of a
scaling assumption and then rechecking this assumption for self-consistency. Consider first the $p_n$ equation, i.e.
Eq.(\ref{eq:p_n}). We substitute $p_n(t)=t^{-1/2}f(nt^{-1/2},t)$ in Eq.(\ref{eq:p_n}) and ask if the resulting
equation allows for a steady state scaling solution as $t\to \infty$, i.e., if the scaling function becomes
explicitly independent of $t$ as $t\to \infty$. It is easy to verify that if $r_1(t)$ decays faster than $t^{-1/2}$, 
such a time-independent scaling solution is possible with $f(x)={x\over 2}\exp(-x^2/4)$. In this case, $l_+(t)=\sum np_n
\approx t^{1/2}\int_0^{\infty}xf(x)dx=\sqrt {\pi t}$ at late times.

Next we consider the $r_n$ equation, i.e., Eq.(\ref{eq:r_n}). Since $p_1=-d{\log N}/dt$, a natural choice would
be to write $r_n(t)=N(t)g(nN(t),t)$. Substituting this in Eq.(\ref{eq:r_n}), we find that in the $t\to \infty$ limit,
the equation allows for a time independent
scaling function, $g(x)=c\exp(-cx)$ ($c$ is a constant), provided $N(t)$ decays faster than $t^{-1/2}$. In this case,
$r_1=N(t)g(0)=cN(t)$. Thus if scaling starts holding beyond some time $t_0$, then $c=r_1(t_0)/N(t_0)$.

Using the results (i) $r_1(t)=r_1(t_0)N(t)/N(t_0)$ and (ii) $l_+(t)=\sqrt {\pi t}$ in the
exact equation Eq.(\ref{Nsol1}), we find
\begin{equation}
{N(t)\over N(t_0)} = \sqrt{{t_0 \over t}} \,{\log(b)\over \log(bt/t_0)}
\label{Nfinal}
\end{equation}
where $\log(b) = {\sqrt{\pi}\over r_1(t_0)\sqrt{t_0}} $.
Substituting this result in the expression for $r_1(t)$, we find
\begin{equation}
r_1(t) = {\sqrt{\pi}\over \sqrt{t} \log(bt/t_0)}.
\label{r1final}
\end{equation}
We now use the late time result Eq.(\ref{Nfinal}) in the exact relation Eq.(\ref{exactP:N}) and find,
\begin{equation}
p_1 = {1\over 2 t} + {1\over t \log(bt/t_0)}.
\label{p1final}
\end{equation}           
From the above expressions, it is clear that both $r_1(t)$ and $N(t)$ decay
faster than $t^{-1/2}$  and hence our scaling solutions are completely self-consistent.

It is easy to see that these IIA results become exact in the zero density limit of the `$+$' phase ($\epsilon\to 0$).
In this limit, the average size of a `$-$' domain is $1/{\epsilon}$ times larger than the average size of a `$+$' domain.
As time increases, the `$+$' domains will
certainly grow in size. But a typical `$+$' domain will disappear (via the absorbing boundary condition) much before encountering
other `$+$' domains, i.e., before feeling the presence of the constraint due to triplets `$+-+$'. The probability of such an event
is of order $O(\epsilon)$. Thus, effectively, the dynamics of the system will proceed via eating up of the `$+$' domains. Hence,
if there is no correlation between domains in the initial condition, the dynamics is not going to generate correlations between them
and hence IIA becomes exact. The picture in this limit is similar to  
the zero temperature dynamics of the $q$ state Potts model in the limit $q\to 1^{+}$\cite{KB}.                                                        
For other volume fractions, it is likely that IIA predicts the correct fixed point picture at late times. This is
confirmed by Monte Carlo simulations of the model.                                                        

To improve efficiency, these simulations were made for a version of the model with 
simultaneous updating. This should not change any of the above conclusions. 
For convenience, we chose initial conditions such that all domains of minority spins had length 1,
while the lengths of majority spin domains were distributed exponentially.  
At each time $t$, all kink positions were written into an array \verb/K/, and only this array 
is used to generate the array \verb/K'/ for the next time step. For each value of $m(0)$ we 
simulated between 20 and 200 lattices of $2^{26}$ sites for $3\times 10^7$ time steps.

\vglue-.3cm 
\begin{figure}
\narrowtext
\epsfig{figure=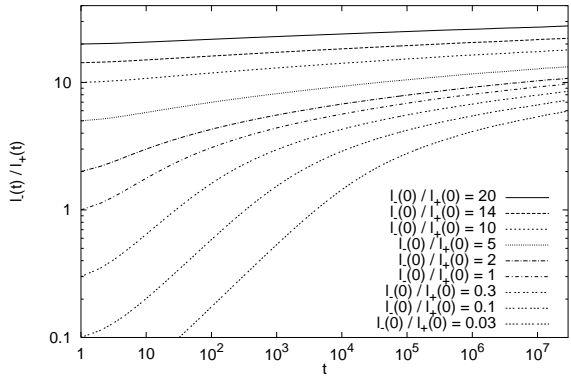,angle=270, width=2.98truein}
\caption{$l_-(t)/l_+(t)$ versus $t$ for runs with different initial magnetizations.}
\end{figure}
\vglue-.4cm

\vglue-.3cm
\begin{figure}
\narrowtext
\epsfig{figure=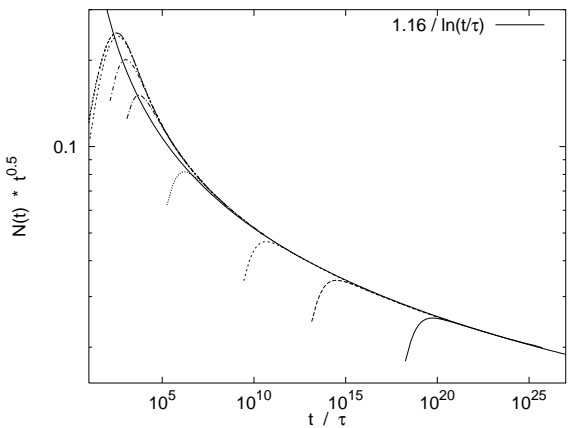,angle=270, width=2.98truein}
\caption{$\sqrt{t}N(t)$ versus $t/\tau[m(0)]$ for the same runs as in Fig.1. The solid line 
is the predition ${\rm const / \log(t/\tau)}$.}
\end{figure}
\vglue-.4cm

Data for $l_-(t)/l_+(t)$, plotted in Fig.1, show the predicted monotonic increase with $t$. This increase is 
logarithmic for $t>t_0$, while it is faster for $t<t_0$. For a detailed comparison with the above theory we 
need to know how $t_0$ (and thus also $b$) depends on $m(0)$. We expect it to be exponential for $m(0)<< 1$,
but this is not sufficient for a detailed analysis. Thus we determine for each $m(0)$ a $\tau$ such that the 
data for $l_-(t)/l_+(t)$, for $\sqrt{t}N(t)$, for $tp_1(t)$ and for $\sqrt{t}r_1(t)$ collapse when plotted 
against $t/\tau$. The fact that a single $\tau[m(0)]$ exists which gives a good data collapse in all four 
plots is highly nontrivial. We just show such plots for $N(t), p_1(t)$ and $r_1(t)$ in Figs.2 and 3. 
We see good agreement with the theoretical predictions. 
In particular, data collapse is excellent (showing that the only memory left from the initial conditions 
is the current value of the magnetization). But the detailed predictions for the scaling function
show substantial corrections which seem however to disappear for $t\to \infty$.

\vglue-.5cm
\begin{figure}
\narrowtext
\epsfig{figure=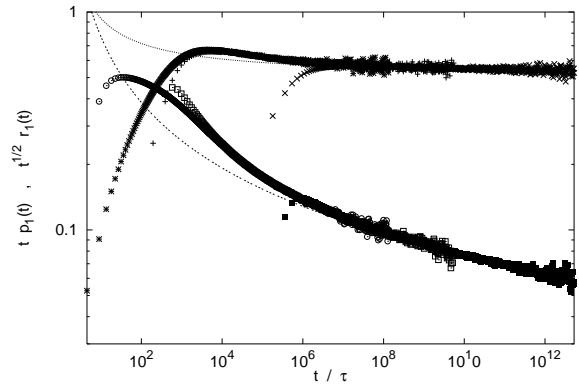,angle=270, width=2.98truein}
\caption{$tp_1(t)$ (top) and $\sqrt{t}r_1(t)$ (bottom) versus $t/\tau[m(0)]$. To avoid overcrowding,
only results for $m(0) = -2/3,0$, and $2/3$ are shown. The dashed curves show the predictions 
$1/2 + 1/\ln(t/\tau)$ and $\sqrt{\pi}/\ln(t/\tau)$.}
\end{figure}
\vglue-.5cm

We thank M. Barma and C. Sire for useful discussions. 
A related reaction diffusion model has recently been studied numerically by Odor and Menyhard
[cond-mat/ 0002199]. We thank G. Odor for communicating his results to us.

\end{multicols}
\end{document}